\documentclass{llncs}
\usepackage[T1]{fontenc}
\usepackage[utf8]{inputenc}
\usepackage{amsmath,amssymb,amsfonts,stmaryrd}
\usepackage{booktabs}
\usepackage{graphicx}
\graphicspath{{figures/}}
\usepackage{hyperref}
\usepackage{mathtools}
\usepackage{xcolor}
\usepackage{xspace}

\newcommand{\R}{\ensuremath{\mathbb{R}}}
\newcommand{\D}{\ensuremath{\mathbb{D}}}
\newcommand{\N}{\ensuremath{\mathbb{N}}}
\newcommand{\ta}{\ensuremath{\tilde{a}}}
\newcommand{\kc}{\ensuremath{k_\text{clock}}}
\newcommand{\km}{\ensuremath{k_\text{AM}}}
\newcommand{\ke}{\ensuremath{k_\text{exp}}}

\newcommand{\AltTextCMSB}[1]{}

\begin{document}

\title{On the Design of an Analog-Dyadic Converter CRN}
\author{Mathieu Hemery\orcidID{0000-0003-2963-7067}}
\institute{Inria Saclay Ile de France, EPI Lifeware, Palaiseau, France}
\maketitle
\thispagestyle{plain}

\begin{abstract}
    The Chemical Reaction Networks (CRN) interpreted through the differential
    semantics, even when restricted to elementary reactions with mass action law
    kinetics, form a Turing-complete language.  This means that any computable real
    function can thus be programmed, and in fact compiled, in an abstract CRN
    that will compute it with an arbitrarily high precision.  In this
    computational framework, the information carriers are the molecular
    concentrations, the required precision is given as input, and the output
    concentration is guaranteed to satisfy the required precision.  On the other
    hand, one can be interested in estimating the derivative of an unknown input
    signal or in reading the concentration value of an input molecular species.
    By nature, such problems can only be approximated with a finite precision.
    Hence, the computation framework proposed previously cannot be applied and
    we need to design and analyze custom CRNs to perform these tasks.  In this
    paper, we present an analog-dyadic converter CRN which takes as input one
    molecular concentration (in $[0, 1]$ but not necessarily computable), and
    produces as output a sequence of ``on'' and ``off'' spikes corresponding to
    some extent to the sequence of bits in the dyadic representation of the
    input concentration.  We provide a detailed analysis of the source of errors
    and their behavior when varying the reactions rate constants. We conclude by
    sketching a possible design for a reader module that takes as input an
    arbitrary concentration and a desired precision and outputs a dyadic
    encoding approximating the value of the concentration with the desired
    precision.
    We leave as an open question to prove the correctness of our construction.
\end{abstract}

\section{Introduction}

Turing machine is a concrete, albeit unpractical, model for the theoretical
study of computers' possibilities. A computer being any machine able to perform
calculations \cite{turing1936computable}. Its interest resides in its simplicity
as it is composed of one or several infinite tapes, a head, a finite set
of symbols to be written on these tapes and a finite set of rules. Despite this
sobriety, the Church-Turing thesis states that Turing machine is one form of the
universal computer in the sense that any function that can be computed by a
computer of any sort can also be computed by it.

A less known ancestor of mechanical computing is the integrating machine of
James Thomson (the brother of Lord Kelvin) \cite{thomson1876integrating} which
later became the differential analyzer and the \emph{General Purpose Analog
Computer} (GPAC) with the work of Shannon \cite{Shannon41} and others. The
central idea in \emph{analog computing} is to represent the mathematical values
of the computation through physical quantities, be it electrical or mechanical
ones, and perform operations on them through the physical manipulation of these
quantities.  The idea, at this time, was mainly to reproduce the behavior of the
model of interest by building a more convenient analog system obeying the same
laws and then perform measurements directly on this ``maquette''. While very
efficient, particularly in terms of energy expenditure, this framework is also
limited in its representative power. Indeed, the solutions that can be
\emph{generated} by such a machine are necessarily constrained by their physical
embodiment. In particular, they are always analytic\cite{BGP17ic}, which seems
to exclude GPAC as a model equivalent to Turing machine.

However, in \cite{GC03} and \cite{BCGH07complexity}, Graça, Bournez \textit{et
al.} realize the \textit{tour de force} to show that GPAC and thus Polynomial
Ordinary Differential Equation (PODE) can have the same expressing power as a
Turing machine and are thus a Turing complete framework. The central idea was to
consider that the function represented by the machine with variables
$\mathbf{x}$ is not the relation: $t \mapsto \mathbf{x}$ -- where the function
is generated by the machine -- but $\mathbf{x}(t=0) \mapsto \lim_{t \rightarrow
\infty} \mathbf{x}(t)$ -- where the function is computed at the limit.  This is
significant because PODE is a purely mathematical notion. If the Church-Turing
thesis is correct, it implies that the computational power of any physical
machine may be defined through a purely mathematical setting.

Building upon these works, our team has shown that through the differential
semantics, chemical reaction networks (CRN) can implement arbitrary
PODE and are thus also a Turing complete framework able to perform any
computation \cite{FLBP17cmsb}. These results drive us to study the theoretical
and practical aspects of CRN seen as a programming language. This leads us to
propose a complete pipeline that compile mathematical functions to
elementary CRN \cite{FHS25cca} and to study the theoretical aspect of the
intermediate operation of quadratization \cite{HFS20cmsb}.

\vspace{1em}

A natural question that is both theoretically interesting and practically
relevant is the method used to read the result of the computation once the
machine has reached the desired precision. It is to shed some light on this
aspect that we propose the present paper.

We present here an analog-dyadic converter that relies entirely on elementary
reactions of chemical species. We take great care to propose an implementation
that uses as few species and reactions as possible while still performing its
function with accuracy and robustness.  The main idea is to receive the
concentration of a species as input and display a sequence of binary spikes that
represents the encoding of the initial concentration as a dyadic number. Hence,
allowing a ``physical observation'' of the initial concentration value with
increasing precision as the number of spikes increases.

From a more theoretical point of view, this constitutes a central and nontrivial
part of the realization of an analog computer, be it with CRN or PODE. Indeed,
the possibility to compute a quantity is of no use if we cannot read the result.
But given the undecidability of the comparison, it is not evident that this task
is actually possible. A careful analysis shows that our device can
only reach a certain precision that depend upon the rate constants and the
initial input. This validity domain exhibits a fractal pattern.  Fortunately, by
tuning the rate constants, it seems possible to expand the validity domain and
reach an arbitrary precision, hence allowing us to build a reader module that
complete the computation module of \cite{FLBP17cmsb}.
To complete the proof of the correctness for the full construction, we would need
a careful analysis of the clock -- and its interrelation with the other parts of
the CRN -- and of the halter module.

\vspace{1em}

Finally, this work is also an attempted step toward the study of the
Hartmanis-Stearns conjecture formulated in 1965 and still pending
\cite{hs65tams}.  This conjecture posits that only the rational or
transcendental numbers can be computed in real time, in the sense that for a
real $x$, there exists a multi-tape Turing machine taking $p$ as input and
returning as output a rational $r$ approximating $x$ with a precision $2^{-p}$
in a time $O(p)$.

Interestingly, the simple CRN:
\begin{equation}
    \emptyset \xrightarrow{5} X,\quad X^2 \xrightarrow{1} \emptyset,
\end{equation}
is such that $X$ converges exponentially fast toward the irrational but
algebraic number $\sqrt{5}$. It is ``simply'' a matter of retrieving this
information.  We will see at the end of this article that this is possible with
a PODE either with bounded concentrations and a time scaling like $o(p^2)$ or
with concentration and time scaling like $o(p)$.  Hence, the Hartmanis-Stearns
conjecture reduces to: ``How can we relate the time of a bounded PODE to the one
of a multi-tape Turing machine?''

\vspace{1em}

The plan of this article is the following. First, we will remind the
main mathematical tools and aspects needed to understand the notion of CRN and
the dyadic representation of numbers, thus setting the notations. We will then
present the functioning of the converter and explain the different elements that
allow it to perform its function in a robust way. After that, we will perform
an-in depth analysis of the main causes of precision loss in order to determine
what is the expected precision and how it scales with the different parameters
of the model. We finally use this analysis to show how to alter this device to
produce a reader module, completing the computation framework of Fages et al.

\section{Notation and computation framework}

\subsection{Chemical Reaction Network}

Chemical Reaction Network is a well-established framework used by chemists
and biologists to represent the interactions of chemical components and study
the various behaviors and properties that emerge from these interactions.

Formally, a CRN is given as a set of species and a set of reactions.  These
reactions are tuples composed of two multisets of species, the reactants
$\mathcal{R}$ and the products $\mathcal{P}$, and a function $f$ indicating the
rate at which the reaction takes place.

Once a CRN is provided, it is possible to interpret it using several semantics:
boolean, stochastic, differential, etc. It is this flexibility that gives CRN
its modeling power, as the same description may be used through a great variety
of tools coming from various domains of science.

In this paper, we will only use the \emph{differential semantics} that consists
in deriving a set of ordinary differential equations (ODE) from the CRN. For
each species, we attribute a variable representing its concentration and build
its derivative by collecting the rates of each reaction in which it participates
and multiply it by its signed stoichiometric coefficient (its index in the
multiset of products minus its index in the one of reactants).

We moreover restrict ourselves to \emph{Mass Action Law} where rate functions
are parameterized by a single positive number $k$ (the rate constant) and are
given by the product of the concentrations of all its reactants multiplied by
the constant $k$. In this case, we obtain Polynomial ODE (PODE). We will use the
standard notation with an arrow going from reactants to products and an upper
index indicating the rate constant if it is different from one:
$$\mathcal{R} \xrightarrow{k} \mathcal{P}.$$

The differential semantics provides a canonical mapping from CRN to ODE that
uniquely associates a set of differential equations to each CRN.
For example, the CRN: $A+B \xrightarrow{k} A+2C$ is associated to the PODE
\begin{equation}
    \dot{A} = 0, \quad
    \dot{B} = -kAB, \quad
    \dot{C} = +2kAB.
\end{equation}

A natural question is then: does there exist an inverse mapping? And the answer
is: yes, but!

Yes, for any ODE $\mathcal{E}$ that uses only elementary mathematical functions,
we can construct a CRN such that a subset of the variables of its associated ODE
through the canonical mapping exaclty admits the same solution as $\mathcal{E}$
\cite{HFS21casc}. For mathematical variables $A$ that may be negative, it is
possible to split them between their positive ($A_p$) and negative ($A_n$) parts
and build a CRN for which the observable $A_p - A_n$ will exactly follow the
initial ODE $\mathcal{E}$ \cite{FLBP17cmsb,HFS21cmsb}. But, this mapping is not
canonical and there is quite a lot of play for the ``implementation'' of a
particular ODE. This is both a curse, because it gives us a lot of arbitrary
choices to make in order to provide a particular CRN, and a boon, because it
allows us to add additional constraints that may bring us desirable properties
upon the constructed CRN.

We study Mass Action Law for two reasons. First, it emerges quite
naturally from collision theory, making them ``more elementary'' than other rate
functions. These are indeed often coarse-grained approximations of the behavior
of more complex Mass Action Law CRN (see \cite{voit2015150} for a historical review
on Mass Action kinetics). More surprisingly, it has been shown that Mass Action
Law CRNs interpreted through the differential semantic are Turing complete in
the following sense\cite{FLBP17cmsb}:
\begin{theorem}
    \label{thm:complete}
    For any computable function $f: \R^k \rightarrow \R$, there exist a finite
    CRN with mass action law kinetics,  over a set of species $S_i$ and a
    polynomial function $p$ such that, if $S_i(t=0) = p(x)$, then the unique solution
    of the differential semantics interpretation of the CRN is such that
    $$\forall t > 1, |S_1(t) - f(x)| < S_2(t), \lim_{t \rightarrow \infty}
    S_2(t) = 0.$$
\end{theorem}

Said otherwise, for any computable function, we can construct a CRN over a finite
number of species that encodes the input upon the initial concentration of its
constituent species, and output the result of the computation as the
concentration of the first species at the end of time while also providing a
bound on the error on the second species.

It is here natural to ask how, at the end of the computation, we can read the
result encoded in the concentration of a chemical species? This is precisely the
question that motivates this paper and for that we will need to introduce the
notion of dyadic number.

\subsection{Dyadic notation}

The set of dyadic numbers denoted $\D$ is the one of numbers that can
be expressed as a fraction whose denominator is a power of two, that is:
$$\D = \left\{x | \exists n,p \in \mathbb{Z} \times \N, x =
\frac{n}{2^p}\right\}.$$
It is a subset of the rational numbers.

By construction, a dyadic number have a natural finite representation which is
the binary equivalent of decimal, for $u$ and $d$ the largest and smallest
needed powers:
$x = \sum_{p=d}^u b_p {2^p},$
where $b_p \in \{0, 1\}$ and of course, $u$ and $d$ can both be positive or
negative.

For example:
\begin{equation}
    \begin{aligned}
        1.8984375 &= 1 + \frac{1}{2} +
               \frac{1}{4} +
               \frac{1}{8} +
               \frac{0}{16} +
               \frac{0}{32} +
               \frac{1}{64} +
               \frac{1}{128} \\
            &= 1.1110011_2
    \end{aligned}
\end{equation}

It is easy to see that $\D$ is dense in $\R$ in the same way that the rationals
or the decimals are.  As for the decimals, a real number always has at least one
(possibly infinite) dyadic representation.  They are thus a convenient set to
represent real numbers in a framework where the number of symbols in the
numeration system is scarce, the same way that binary allows representing
integers with only two symbols.


We now have all the elements to present the basic structure of our converter.

\section{The analog-dyadic converter}

\begin{figure}
    \centering
    \includegraphics[width=0.65\textwidth]{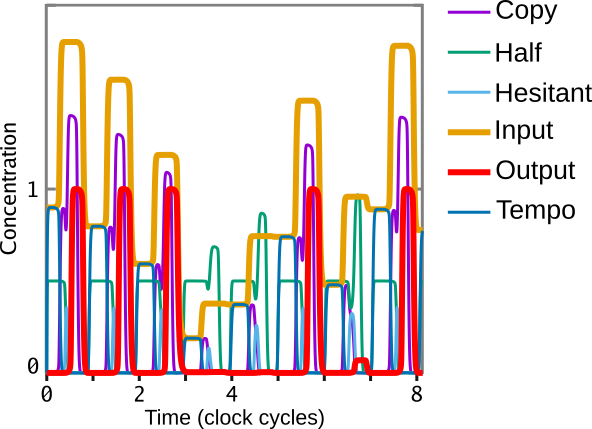}
    \caption{Time course of the dyadic converter through $8$ complete cycles
    of the clock. During the $k$th cycle, the $Output$ species (in bold red)
    spikes or not according to the comparison of the $Input$ species (in bold
    orange) to a
    threshold species. Here we can see that the output is $11100101$.
    Note that this is different from the expected result as,
    $\text{Input}(t=0) = 0.9 = 0.11100110\ldots_2$. The $6$ first bits are
    correct, but we suffer from an error on the $7$th one. The cause of these
    errors, and how to mitigate them, will be discussed in this paper.}
    \AltTextCMSB{Time course of the Copy, Half, Hesitant, Input, Output and
    Tempo species of the dyadic converter through $8$ cycles of the clock. The
Output species in red exhibit a clear binary pattern with 3 spikes, two cycles
without spike then a spike and an aborted spike and finally a true spike. The
concentrations of all species being bound between $0$ and $2$.}
\label{fig:09_conversion}
\end{figure}

The purpose of the dyadic converter is, starting from a species, the
concentration of which is between $0$ and $1$, to ``write'' the dyadic
representation of this initial concentration as a sequence of binary spikes in
the concentration of one species (here named ``Output''). 
An example of this behavior is shown in Figure~\ref{fig:09_conversion}.

The main idea is simply to follow the same steps as those of the typical
algorithm used to convert a number to its dyadic representation.
In a nutshell, we first compare the current value of the
input to one half and write a one or a zero accordingly. Then, in order to pass
to the next bit, we need to zoom in. This can be done by doubling the input and
subtracting one if necessary to keep it between $0$ and $1$. Then we loop.

The reactions of the dyadic converter are a bit more involved as they include
ancillary species to perform the function with the constraints of chemical
implementation. The pseudocode along with its CRN implementation is given
in table~\ref{tab:pseudocode}.

\begin{table}
    \centering
    \begin{tabular}{ll|c}
        & \large{Pseudocode} & \large{CRN} \\
        \hline
        Copy & $Copy := Input$ & $Input \rightarrow Input + Tempo$ \\
        \&     & $Input := 2*Input$ & $Tempo \rightarrow \emptyset$ \\
        Doubling & & $Tempo \rightarrow Copy + Input$ \\
        \hline
                 & $Half := \frac{1}{2}$ & $\emptyset \rightarrow Half, \quad
                 Half \xrightarrow{2} \emptyset$ \\
        Comparison & $Half := 0 \text{ if } Half < Copy$ & $Copy + Half \rightarrow 2 Hesitant$ \\
                 & $Copy := 0 \text{ if } Half > Copy$ & $Hesitant + Half \rightarrow 2 Half$ \\
                 &                                     & $Hesitant + Copy \rightarrow 2 Copy$\\
        Spiking  & $Output := 1 \text{ if } Copy$ & $Copy + Inactive
        \leftrightarrow Copy + Output$ \\ 
        \hline
        Subtraction & $Input := Input - Output$ & $Output + Input \rightarrow
        Inactive$
    \end{tabular}
    \caption{Description of the operations performed by the dyadic converter module
    given with a pseudocode representation along with the CRN implementation.
For clarity, the clock and its regulation are not represented here. The exact
model is given in (\ref{eq:full_crn}).}
    \label{tab:pseudocode}
\end{table}

\subsection{Overview}

Our conversion module is composed of five interacting parts: a clock, a copy, a
comparison, a subtraction and a cleaning mechanism.

The purpose of the clock is to play the role of a musical director. It
coordinates all the other reactions as they should be
activated in order to correctly perform the desired behavior.
This is achieved by having a set of species that alternate active and passive
phases in a predefined order and use them to catalyze all the reactions of the other
parts.

The copy reactions allow us two different effects. First, it creates a copy of
the input that can be destroyed during the comparison step without losing the
value of the input that should be passed (up to some modifications) from one
cycle to the next. It also allows us to double the value of the input.

The comparison scheme is based upon the classical mechanism of \emph{approximate
majority} \cite{CC12sr}. It compares with good precision the value of the Copy
to a predefined threshold, here: one half. While a more simple comparison could
have been implemented with a simple bidegradation, the approximate majority
scheme is here preferred for its robustness and the strength of its signal.
Indeed, at the end of the computation the reminding species is always higher
than one half, hence ensuring an easy reading. Of course, this mechanism is not
perfect, and it takes time to discriminate, especially between close
concentrations, a phenomenon that is the core of section \ref{sec:am_detail}.

The subtraction reaction is a simple bidegradation. It relies on an output
species that is activated if and only if the input is higher than one half based
on the result of the previous comparison mechanism. This has two benefits:
first, it creates a robust output signal that is simply $0$ or $1$. Second, as
it goes to $1$ independently of its previous value, it provides us with
the unitary value that should be subtracted from the input to start the next
cycle of the dyadic conversion as explained in the pseudocode.

The cleaning reactions finally simply remove the intermediate species in order to start
the next cycle with null concentration. This step is facultative
as these species are set by the different reactions to their desired values
independently of their initial concentrations.
However, cleaning them ease the analysis of the model.

We will now take a bit more time to present in detail the two more involved
elements of the CRN, namely the clock and the comparison module.

\subsection{Clock}

It is now known for some time that you can easily generate a sine and cosine
with 4 species and 6 reactions~\cite{FLBP17cmsb}. In this CRN, the 4 species are
the positive and negative parts of the trigonometric functions while the
reactions are, for four of them, the catalytic production of the next species,
the two others being bidegradations of the positive and negative parts.

\begin{figure}
    \centering
    \includegraphics[width=0.7\textwidth]{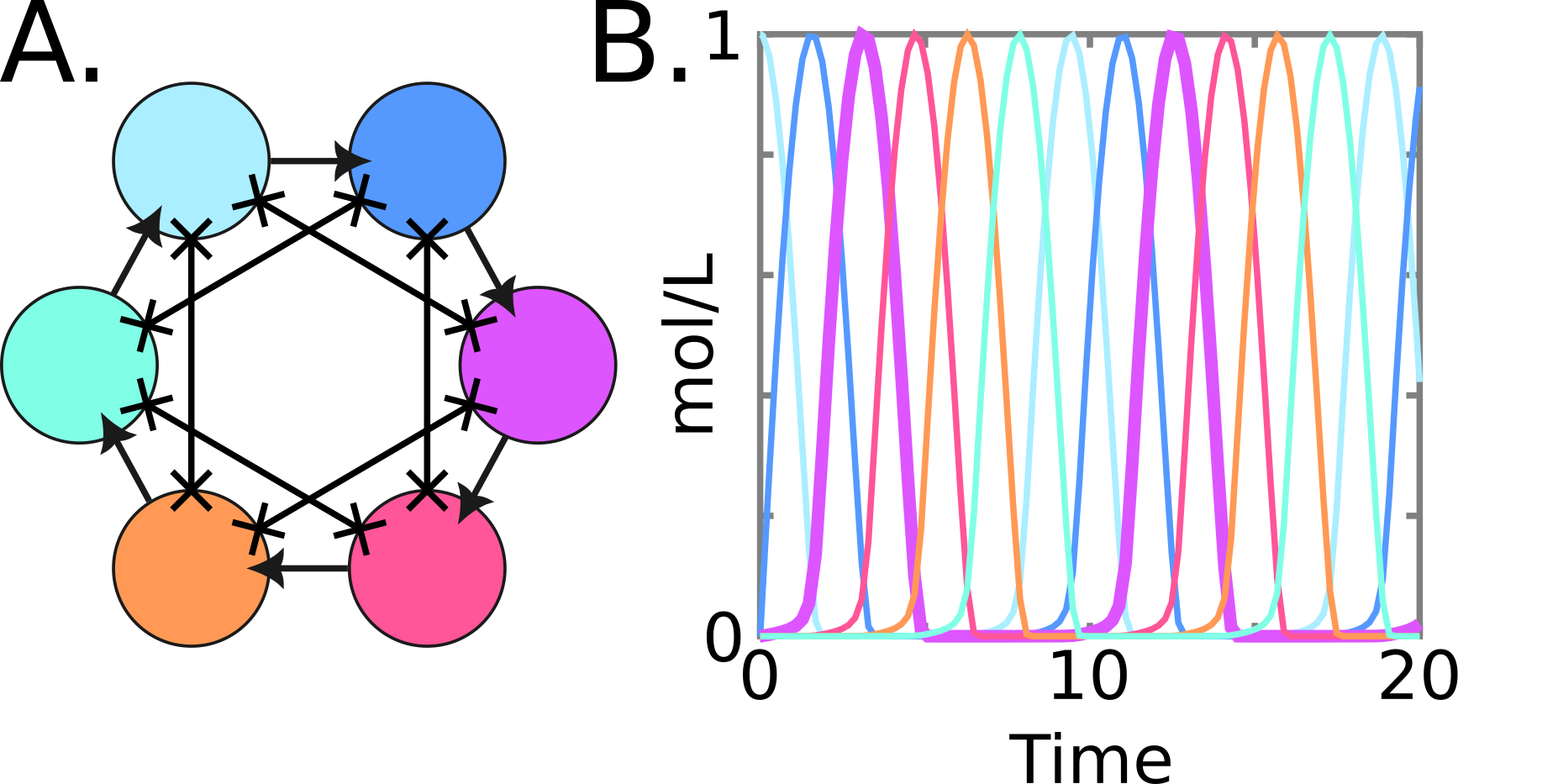}
    \caption{
    \textbf{A.} Schematic representation of a six-step clock. Each species
    activates the next one and participates in a bidegradation (here depicted
    as a two crossed connector) with its second neighbor.
    \textbf{B.} The time course of the CRN presented in panel A. with
    corresponding colors. Activations have rates $\kc = 1$ while
    degradations have 
    rates $k_f = 1000$. At $t=0$ one of the species (here light blue) has an
    initial concentration of $1$. The violet species has been highlighted only to ease the
    readability of the behavior of a single species.
    }
    \AltTextCMSB{In pannel A, six species are depicted in a circle, each one
    activates its successors and participates in a bidegradation with the species
two neighbors away. In pannel B the time course of the clock species, one
species in bold clearly illustrates the pattern of sinusoidal active phase and
null dormant phase.}
    \label{fig:clock}
\end{figure}

Surprisingly and as presented in Figure~\ref{fig:clock}, this is not limited to
4 species. For any $n \geq 4$, the CRN:
\begin{equation}
    T_i \xrightarrow{\kc} T_i + T_{i+1}, \quad
    T_i + T_{i+2} \xrightarrow{k_f} \emptyset,
\end{equation}
where the indices $i \in [1, n]$ and should be understood modulo $n$ and the
rate $k_f$ is high enough to ensure that bidegrations strongly dominate other
reactions, also
exhibits an oscillatory behavior. Even more beautiful, it gives us an $n$-clock where
all the species alternates between an "active" sinusoidal behavior corresponding
to one "bump" and a "dormant" null behavior where it simply waits to be activated
again.

By using each of these species as a catalyst for other reactions, we can
associate a part of the CRN of table~\ref{tab:pseudocode} to a particular time
of the $n$-clock, the same way
the circadian clock regulates the activity of pathways in cells.

For a computer scientist, our clock is thus of a surprising kind as one complete
cycle of the clock does not correspond to one elementary operation of the system
-- as it would in a traditional computer with a clock used for synchronizing the
different parts -- but to a full cycle of operations leading to the computation
of one bit of the dyadic encoding.

\vspace{1em}

For a perfect computation, the steps of the conversion should be performed
sequentially and independently.
The main problem is that our clock is such that two consecutive species, $T_i$
and $T_{i+1}$ display a strong overlap, as can be seen in Figure~\ref{fig:clock}. Even the overlap between $T_i$ and $T_{i+2}$
may be too important when strong guarantees are needed: typically when the two
steps are likely to create a positive feedback loop that may lead to an
exponential divergence as is the case for
the two reactions of the copy mechanism: $Input \rightarrow Input+Tempo$ and $Tempo
\rightarrow Copy + Input$. To avoid that, we
can always let more steps of the clock pass between
such reactions. We may also increase the rate constant $k_f$
as it helps to soothe this unwanted overlap.

As this work is also a proof of principle for a short device, we
try to minimize the use of such insulation mechanisms to keep our device as
streamlined as possible. In the context of a formal derivation of the
theoretical possibility of CRN computation, we would take $n=80$ and associate
the different step to $T_{10 \cdot i}$ instead of $T_i$ to ensure strong
insulation.

\subsection{Thresholding with Approximate Majority}
\label{sec:am}

The comparison relies first on the production of a threshold species (H for Half):
\begin{equation}
    \label{CRN:half}
    \emptyset \xrightarrow{k_n} H, \quad H \xrightarrow{k_d} \emptyset,
\end{equation}
that will converge to the ratio: $\frac{k_n}{k_d}$. In our case, we use $k_d = 2
k_n$ to obtain a one half ratio. This step is perform as an initialization and
realized during the first step of the clock.

Then during step $5$, an approximate majority scheme is performed where
the species of interest $A$\footnote{To avoid cluttering the notation of this
    section, we use
    $A$ and $a$ for the study of the approximate majority in its own. In the
    full CRN, the species to be compared will be $Copy$. Similarly, the
undecided $U$ species is named Hesitant in the full CRN.} is compared to the $H$
species through the formation of an undecided $U$ species:
\begin{equation}
    \label{CRN:AM}
    A + H \xrightarrow{k_1} 2 U, \quad
    U + A \xrightarrow{k_2} 2 A, \quad
    U + H \xrightarrow{k_2} 2 H.
\end{equation}
In model~(\ref{eq:full_crn}), we set $k_1 = k_2 = \km = 10$. We keep them
different here for the analysis.

The differential semantics of this reaction system gives us the following ODE:
\begin{equation}
    \begin{aligned}
    \dot{A} &= -k_1 A H + k_2 U A, \\
    \dot{H} &= -k_1 A H + k_2 U H, \\
    \dot{U} &= 2 k_1 A H - k_2 U (A+H).
    \end{aligned}
\end{equation}

The approximate majority is well known to converge to a steady state where the
species that was initially the more abundant between $A$ and $H$ will be the
only one remaining with a concentration equal to the sum of the two initial ones.
That is, we have when $t \rightarrow \infty$:
\begin{equation}
    \begin{cases}
        A = A_0 + H_0,\quad H = 0 &\text{if } A_0 > H_0, \\
        A = A_0,\quad H = H_0 &\text{if } A_0 = H_0, \\
        A = 0,\quad H = A_0 + H_0 &\text{if } A_0 < H_0. \\
    \end{cases}
\end{equation}


\vspace{1em}

Using this scheme, we can thus test if the initial concentration of $A$ was below
or above the threshold $H$ just by testing its presence at the end of the
process. This supposes, that the approximate majority dynamic had enough
time to converge. Or at least that, if it should be $0$, the eventual traces of
the $A$ species are low enough to remain undetected. This question will be
treated in length in section~\ref{sec:am_detail}.

Now that we have presented all the elements, let us see how they perform
together.

\subsection{Dyadic converter}

The full CRN of our converter is given by:
\begin{small}
\begin{equation}\label{eq:full_crn}\begin{aligned}
    \forall i\in[1,8] T_i &\xrightarrow{\kc} T_i + T_{i+1}, &\quad
    \forall i\in[1,8] T_i + T_{i+2} &\xrightarrow{k_f} \emptyset, \\
    Input + T_1 &\xrightarrow{\ke} Input + T_1 + Tempo, &\quad
    Tempo + T_1 &\xrightarrow{\ke} T_1, \\
    T_1 &\xrightarrow{\ke} T_1 + Half, &\quad
    T_1 + Half &\xrightarrow{2 \ke} T_1, \\
    T_4 + Tempo &\xrightarrow{\ke} T_4 + Input + Copy, &\quad
    T_5 + Copy + Half &\xrightarrow{\km} T_5 + 2 \cdot Hesitant, \\
    T_5 + Hesitant + Half &\xrightarrow{\km} T_5 + 2 \cdot Half, &\quad
    T_5 + Hesitant + Copy &\xrightarrow{\km} T_5 + 2 \cdot Copy, \\
    T_6 + Copy + Inactive &\xrightarrow{\ke} T_6 + Copy + Output, &\quad
    T_7 + Tempo &\xrightarrow{\ke} T_7, \\
    T_7 + Copy &\xrightarrow{\ke} T_7, &\quad
    T_7 + Half &\xrightarrow{\ke} T_7, \\
    T_8 + Input + Output &\xrightarrow{\ke} T_8 + Inactive.
\end{aligned}\end{equation}
\end{small}
With the initial condition: $T_1 = 1, Inactive = 1$ and of course $Input = I_0$.
For this paper, we set: $\kc = 1, \km = \ke = 10, k_f = 1000$.

\begin{figure}
    \centering
    \includegraphics[width=0.9\textwidth]{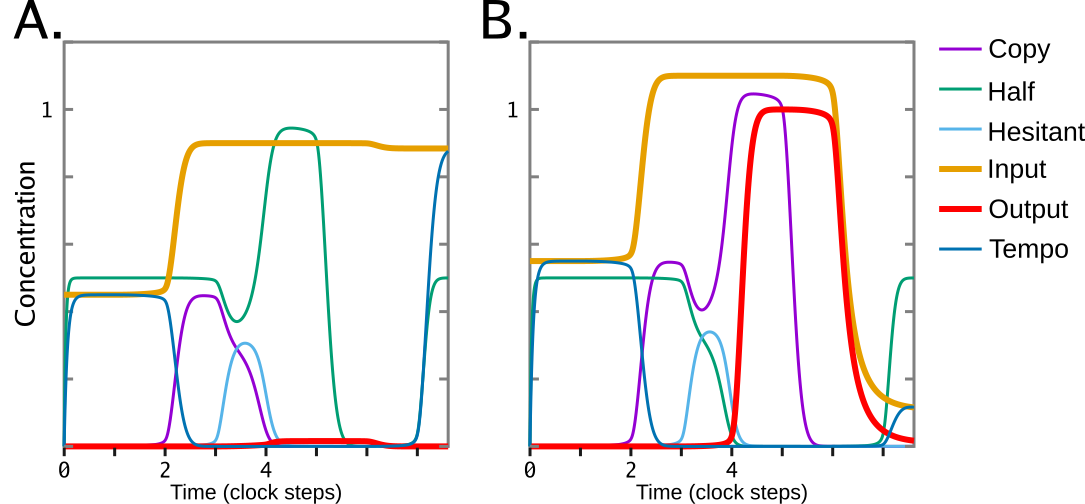}
    \caption{Numerical integration of one full cycle of the clock for two
    different initial conditions. One where the $Input$ species is slightly below
(\textbf{A.}) the $\frac{1}{2}$ threshold and one where it is slightly above
(\textbf{B.}).}
    \AltTextCMSB{Time course of the Copy, Half, Hesitant, Input, Output and
    Tempo species of the dyadic converter through $1$ cycles of the clock. In
pannel A, the initial input is set to $0.45$ and ends up at $0.9$, the Half
species goes to $0.5$ and then $0.95$. There is no spike on the Output species. In
pannel B, the initial input is set to $0.55$ then reach $1.1$ and ends up at $0.1$, the Half
species goes to $0.5$ and plummets to $0$. There is a clear spike in the Output
species between clock steps $4$ and $6$. }
    \label{fig:full_cycle}
\end{figure}

In Figure~\ref{fig:full_cycle}, we illustrate the behavior of the model through
one full cycle of the clock depending on whether $I_0$, the initial value of the
input,
is below or above the
threshold of $\frac{1}{2}$ and thus whether the output species should spike or
not. We can see that at the end of step $4$, the $Copy$ species now has the
value $I_0$ and $Input$ itself has doubled its value. Then
the $Copy$ is compared to $Half$ and thus set back to $0$ in the first case,
while it goes to $\frac{1}{2}+I_0$ in the second one. The presence or
absence of $Copy$ at this time activates (or not) the production of $Output$ to
its target value of $1$. It is this unitary (or null) value which is then
subtracted to the current value of $Input$ to set it to its value for the next
iteration of the clock, that is either $2 I_0$ or $2 I_0 - 1$.

We just show that our simple CRN is able to display the desired behavior by
implementing the pseudocode of table~\ref{tab:pseudocode} to output at least
the first bit of the dyadic encoding of its input. We have also seen that this
network is susceptible of making mistakes. After a certain time, the
initial precision is lost and the bits no longer correspond to the exact
representation. We call \emph{precision} the number of correct bits that the CRN
is able to output. This quantity depends on the parameters of the CRN and the
value of the initial input.

We will now take time to determine what are the main sources of such
discrepancies and how we can control them. We will thus determine the influence
of the parameters of the CRN on these errors and give some theoretical bounds
upon the precision of our conversion device.

In particular, we will estimate the validity domain of our device. Indeed, we
have seen that the precision we can achieve depends on both the input and the rate
constants. For a given set of parameters ($\kc, \ke, \km, k_f$), it is this
attainable precision as a function of the input that we want to determine.

\section{Detailed analysis}

Our analysis of the errors relies on three properties:
\begin{itemize}
    \item Converging exponential have had enough time to reach a precision
        $\epsilon$ with respect to their target value during one step of the
        clock (see section~\ref{sec:epsilon} but also appendix \ref{app:cosine}).
    \item The different steps of the clock are well separated. This avoids any
        leakage between successive steps. While this is only approximately the case for our
        8-clock example, we can increase the size of the clock to ensure
        perfect insulation.
    \item The Approximate Majority mechanism had enough time to
        discriminate between its two constituents up to a precision
        $\delta^\star$ (see
        section~\ref{sec:am_detail} and appendix \ref{app:AM}).
\end{itemize}

Conceptually, the two first properties may be thought as the use of a perfect
clock where the pieces of cosine have been replaced by rectangular signals of
amplitude $1$ and duration $\tau$. Hence, the different parts of the module are
activated at full intensity for a constant duration before being switched off.
For this reason, we rely on the parameter $\tau$ as a convenient parameter to
describe
the clock. As shown in appendix \ref{app:cosine}, we have the relation $\tau
= \frac{2}{\kc}$.

For the last property, we will see that it creates a constraint on the different
rate constants of the reactions.

\subsection{Premature halting}\label{sec:epsilon}

The main culprit in terms of error is the premature halting of the reactions before
they reach their asymptotic steady states. Note that this error is
unavoidable as the exact computation is only performed in infinite time due to
the framework of ``computation at the limit''.

Most differential equations involved in the dyadic converter harbor
converging exponential with a rate $\ke$ as solutions.
Knowing this allows us to compute the bounds on the value of the different
species during the time evolution of one cycle of our converter module as a
function of the input value $I_0$. This allows us to bound $I_f$ the value of
$Input$ at the end of the clock cycle (the full detail of this derivation is
given in appendix \ref{app:precision}):
\begin{equation}
    \begin{cases}
        I_f \in [2I_0 - 3\epsilon, 2I_0] &\text{if } I_0 < \frac{1}{2}, \\
        I_f \in [2I_0 - 1 - 2\epsilon, 2I_0 - 1 + \epsilon] &\text{if } I_0 >
        \frac{1}{2}.
    \end{cases}
\end{equation}

An error of order $3 \epsilon$ after each cycle means that the value of the
Input after $n$ cycles has accumulated an error like:
$$e_n = 3 \left( 2^n - 1 \right) \epsilon.$$
Thus if we still want a good precision at step $p$, this means that we want $3
\cdot 2^p \epsilon \ll 1$ and for this, we have to impose: $\ke \tau \gg p$.

Reaching the $p$th cycle asks for a time $T(p) \propto \tau p$, and as we have
$\tau \propto p$ by our previous computation, we conclude that reaching a given
precision asks for a time that scales quadratically:
\begin{equation}
    \label{eq:quadratic_time}
    T(p) \propto \frac{p^2}{\ke}.
\end{equation}

However, the premature halting of the converging exponential is not the only
cause of errors that can be perpetrated by our converter. The other main cause
of errors being the incapacity to distinguish between two similar
concentrations when the input and threshold species are too close, that is, when
the input is near one half and this is what we discuss in the next section.

\subsection{Comparison}\label{sec:am_detail}

Analyzing the ODEs of the approximate majority \ref{CRN:AM}, (see appendix
\ref{app:AM} for the full
derivation), we deduce that the system does not have the time to reach a proper
convergence when the initial value is too close from one half. Introducing
$\delta \geq 0$
as $I_0 = \frac{1}{2} \pm \delta$ we can derive that when $k_1 \gg k_2$, there
exists a threshold $\delta^\star$ given by:
\begin{equation}\label{eq:delta_star}
    \log (\delta^\star) = -\frac{\log (4 r^2)}{1+\sqrt{2}} - \frac{k_2 \tau}{1+\sqrt{2}}
\end{equation}
such that starting with a lower value for $\delta$ indicates that the
approximate majority mechanism will
at time $\tau$ be in a state where the species that should be close to $0$ is
actually higher than $r$. For ensuring the
bounds of the previous section, we need $r < \frac{\epsilon}{\tau \ke}$.
Note that this relation introduce a dependency between the error sources,
lowering $\epsilon$ also lower $r$ which leads to higher $\delta^\star$.

When $\delta < \delta^\star$ the comparison mechanism will not have the time to
converge properly, this removes $[\frac{1}{2}-\delta^\star,
\frac{1}{2}+\delta^\star]$ from the validity domain for precision $1$. During
the next cycle, the same phenomenon appears, creating two new forbidden
intervals:
$[\frac{1}{4}-\frac{\delta^\star}{2}, \frac{1}{4}+\frac{\delta^\star}{2}]$ and
$[\frac{3}{4}-\frac{\delta^\star}{2}, \frac{3}{4}+\frac{\delta^\star}{2}]$
for the precision $2$.
And the same process repeats at each clock cycle, creating
the fractal pattern of the validity domain represented in black in
Figure~\ref{fig:precision} A. For this picture, we choose:
$\delta^\star = e^{-\frac{10 \cdot 1}{3}} \simeq 0.036.$

\vspace{1em}

To conclude on the behavior of the approximate majority scheme, it has to be
said that with a finite number of molecules, the stochastic aspect of this CRN
would also impose some limit.
We would have to set a limit on fault tolerance: ``I want my
discrimination to be correct in at least $99\%$ of the cases'' and this would 
translate into a new limit $\delta_\text{stoch}$ with a behavior similar to
$\delta^\star$, effectively setting a limit to how far we could lower
$\delta^\star$.

Now let us check if these theoretical considerations are valid once we start
making numerical simulations of our device.

\subsection{In practice}

To evaluate the actual quality of the response, we perform extensive numerical
simulations, always using model (\ref{eq:full_crn}) for
$1000$ different values of the initial input in the range $[0,1]$.
We then automatically extract the presence or absence of the $10$ first
spikes\footnote{For this analysis, we consider a spike to be present if the
Output species spikes above a threshold of $\frac{1}{3}$ thus making a
pessimist assumption about the ability of the reader to discriminate
intermediate value. It could be interesting to add an \emph{unclear} state for
these intermediate values and see how this allows to catch the precision loss of
the module.} recovering to the effective encoding (which may be different from
the mathematical exact one).

\begin{figure}
    \centering
    \includegraphics[width=0.9\textwidth]{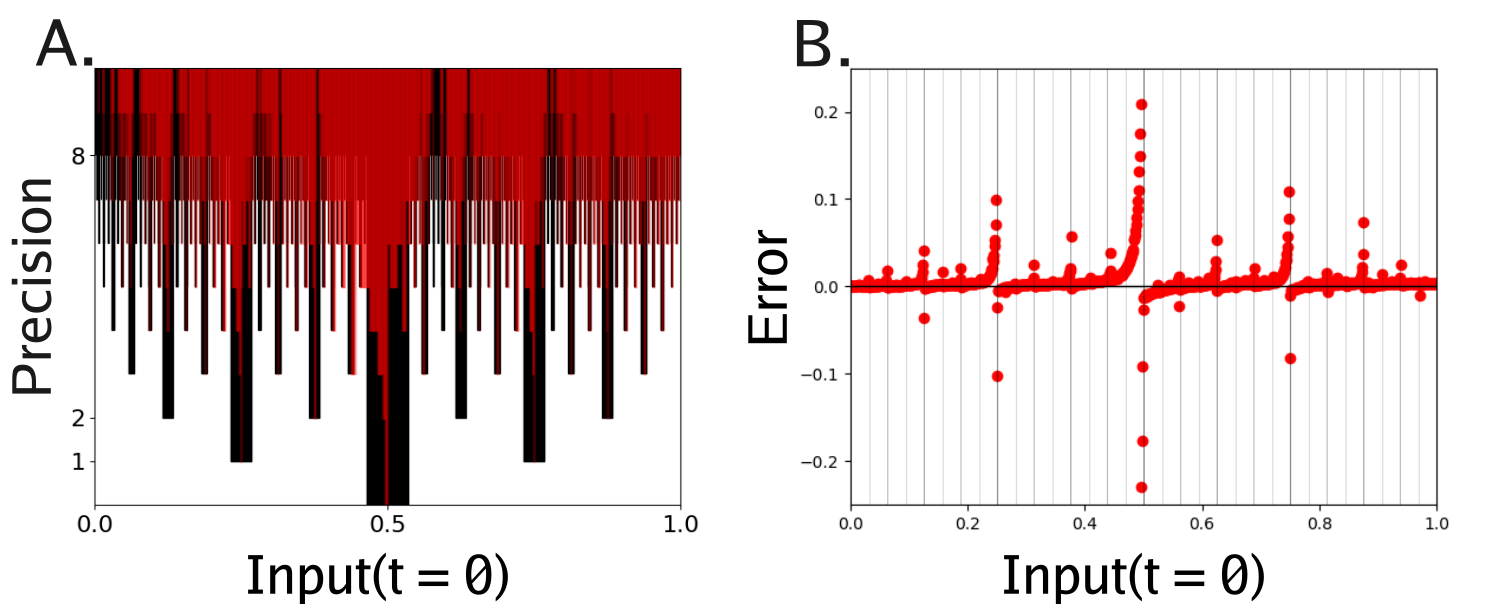}
    \caption{
        Analysis of the actual behavior of the converter module for $\kc = 1$
        and $\ke = \km = 10$ and $1000$ different values of the input between
        $0$ and $1$. The
        system is then numerically integrated, and the spikes are automatically
        detected to determine the system response over $10$ full cycles
        of the clock.
        \textbf{A.} Comparison between the theoretical precision of the
    approximate majority analysis (in black) and the actual precision of the
full CRN as measured after numerical integration (in red). \textbf{B.} \emph{Error} of the dyadic converter as a function of the initial input.}
    \AltTextCMSB{Pannel A shows in red, the precision of the converter described
    in the text as a function of the input. And in black the fractal pattern
of the theoretical limit of precision due to the behavior of the approximate
majority mechanism. The red bars are essentially inside of the black region,
indicating that for this parameter set, the main errors come from the
approximate majority mechanism.
Pannel B shows the error as a function of the input. We clearly have hyperbolic
patterns where the error increases, reaches the asymptote and then goes back from
a low value toward $0$ just after the asymptote. This pattern is manifest at one half
and repeats for each dyadic number with a low denominator in a fractal way. }
    \label{fig:precision}
\end{figure}

We propose two different measures to determine the precision
of our model:
\begin{itemize}
    \item The \emph{precision} defined as the first bit where the model makes an
        error (depicted in red in fig.\ref{fig:precision}A.),
\item The \emph{error} defined as the signed difference between the actual
    number and the number encoded by the dyadic representation output by the
    model (see fig.\ref{fig:precision}B).
\end{itemize}
For example, in the simulation presented in Figure~\ref{fig:09_conversion}, the input is
$0.9 = 0.1110011001100\ldots_2$ and the output of the model is
$0.11100101\ldots_2 = 0.8945$. In this case, the precision is $7$ and the
error is $-0.0055$.

\emph{Precision} and \emph{error} are two different measures and a low precision
is only a piece of evidence for a high error. Think of $0.500002$ and $0.49995$ which
differ very slightly despite a difference as early as the first digit. It is
thus important for us to look at both.

We can see on the panel \textbf{A.} that the
actual precision is more subtle than the theoretical picture presented in
black. This is because of the presence of errors due to premature halting but
also due to our choice to present a streamlined CRN with as few reactions
as possible and to take low values for the rate constants. This allows
us to show explicitly the behavior of our converter in its different error
regimes.

First, it needs to be emphasized that both panels in Figure~\ref{fig:precision}
exhibit the fractal pattern discussed above. Moreover and as expected, the
precision is essentially bounded by the results of our AM analysis.
We also see that the effect of premature halting is
more important for high values of the input, explaining the lower precision near
$1$. (This is also true for error, but is difficult to see due to the scale of
the figure.)

Let us now discuss briefly the behaviors of the CRN that were not covered by our
previous analysis.
To derive our computation, we tried to estimate a lower bound on precision
and rely on two hypotheses: first, a pessimistic approach to the error, since as
soon as the parameter $\delta$ was below the critical threshold, we consider
the system to be faulty, and second, an exact value for the quantity to be
compared. Sadly, both assumptions are false.

Manifestly, errors are not ineluctable inside this black region, but being too
close to $\frac{1}{2}$ indeed incurs a loss of precision as the system is
halted before it reaches consensus.  First, we can see that the error is the
worst slightly below one half. This comes from the fact that the Half species
we compare to suffer from the premature convergence effect discussed in
\ref{sec:epsilon}. For this set of parameters, the actual threshold seems to be around
$0.496$.

The way the approximate majority errors affect the result is also interesting.
Indeed, due to the continuous nature of our computation, around one half, it is
possible to see ``aborted'' spikes of the Output that are neither $0$ nor $1$.
This kind of behavior can be seen around time $80$ in
Figure~\ref{fig:09_conversion}. Through the subtraction part, this error 
propagates to the Input species meaning that the next
cycle start with an incorrect input.

Typically, just after the one half threshold, we would expect the input species
to be close to zero at the beginning of the second clock cycle. If this was the
case, we would only get zeros after that, and so, a low error. But as can be
seen on panel \textbf{B.}, it is not the case.
This creates the hyperbolic pattern that appears around one half and, by the
fractal pattern mentioned before, around all dyadic numbers with a low denominator as
indicated by the black vertical lines.

\vspace{1em}

Let us now talk about the scaling of these errors with the different rates of
our model. Amusingly, the responses of the system to a tuning of the different rates
are not independent. While we would have thought that a large increase in $\ke$
without modifying $\km$ would reduce the source of error to only the problematic
cases around one half -- and its fractal successors -- it is not what actually
happens. Indeed, a high $\ke$ makes the spiking reaction so sensitive, that even
a residual fraction of the Copy species is detected and trigger the spike as 
revealed in equation~\ref{eq:delta_star}.

Conversely, increasing only $\km$ effectively narrows the problematic
fractal region by decreasing $\delta^\star$ until the precision is essentially
limited by the problem of premature halting.
Hence, effectively increasing the precision necessitates to increase $\ke$ while keeping
$\km \gg \ke$.

All this can be used to build a \emph{reader} module, completing the
computation module of Fages et al.

\subsection{Building a reader module}

Theorem \ref{thm:complete} ensures that for any computable function $f$, we can
build a CRN on species $\mathbf{S}$ such that for any $x$, the first species
converges to $S_1 \rightarrow f(x)$ and the second species, which is positive,
decreases and converges to $0$, controls the error:
$|S_1 - f(x)| < S_2$.
For a desired precision $\epsilon$ we can wait until $S_2(t^\star) =
\frac{\epsilon}{2}$ and return any value past this point in time. We choose to
stop the computing device at a particular time $t
\geq t^\star$ and denote $\tilde{f}(x) = S_1(t)$. Then you need to start the
reading device. A prototype of linker CRN is presented in
appendix~\ref{app:linker} that allows to let the computing CRN converge before
halting it and starting the reader module.

\begin{figure}
    \begin{center}
    \includegraphics[width=0.6\textwidth]{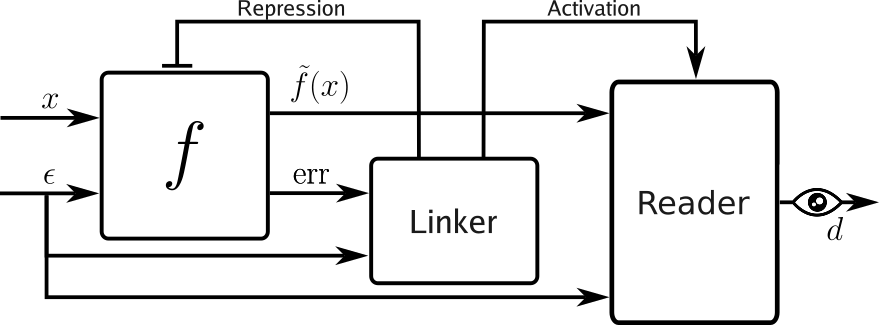}
    \end{center}
    \caption{Schematic figure of a complete analog machine to compute the
        function $f$. The first module takes as input $x$ and the desired
        precision $\epsilon$ and outputs a (possibly not computable) value
        $\tilde{f}(x)$ while also providing its current estimation of the error
        on err. A linker module halts the computation once the precision is
        reached
        and activates a reader module that takes both the output of the previous
        module and the desired precision and returns a physically observable
    signal which allows us to construct a dyadic $d$ encoding the result with
desired precision.}
    \AltTextCMSB{The picture displays three boxes representing the three CRN for
    computation, linker and reader, along with arrows indicating their input,
output and how the linker both represses the computer and activates the reader once
the desired precision is reached.}:w

    \label{fig:reader}
\end{figure}

The difficulty is that $\tilde{f}(x)$ is \textit{a priori}, an ordinary real.
There is no reason why it should be computable. We only know that it approaches
the result with the aforementioned precision.  Starting from the converter CRN,
we can build a reader module that takes this value as input along with the
precision $\epsilon$ and produces an observable behavior allowing the user to
construct a dyadic number $d$ such that $|d - \tilde{f}(x)| \leq \frac{\epsilon}{2}$.
Hence, we have with $d$ a dyadic approaching $f(x)$ with the desired precision
$\epsilon$ as outlined in Figure~\ref{fig:reader}.

To reach this precision, we need to have both $\epsilon \geq e^{-\ke \tau}$ and
$\epsilon \geq e^\frac{- \km \tau}{1+\sqrt{2}}$. We suppose that we are given the precision as
a species with a concentration $p$ such that $\epsilon = e^{-p}$. We thus need $\km > \frac{3
p}{\tau}$ and $\ke > \frac{p}{\tau}$. To allow the CRN to adapt automatically
to the desired precision, the idea is to
introduce a catalyst species to tune the rates at which the reactions take place,
in order to obtain a satisfying precision. In practice,
we may simply use $p$ as the tuning species.

The approximate majority part, for example, would thus become:
\begin{small}
\begin{equation}
\begin{aligned}
    T_5 + P + Copy + Half &\xrightarrow{3 \km} T_5 + P + 2 \cdot Hesitant, \\
    T_5 + P + Hesitant + Half &\xrightarrow{3 \km} T_5 + P + 2 \cdot Half, \\
    T_5 + P + Hesitant + Copy &\xrightarrow{3 \km} T_5 + P + 2 \cdot Copy.
\end{aligned}
\end{equation}
\end{small}
And similarly for all the reactions involving $\ke$ and $\km$.

To conclude this section on the reader module. The attentive reader may see a contradiction
between the quadratic time discussed in sec.~\ref{sec:epsilon}, and the fact that we can
adapt to arbitrary precision without modifying the clock in this section (hence
achieving a $o(p)$ time). The
crucial difference lies in the bounds of the ODE. When adjusting the time of the
clock, all the solutions of the ODE are bounded by a constant ($2$ in practice).
While in the reader module, the concentration of the precision species increases with the
precision (quite obviously). Howerver, we should also stress that it is unclear
how exactly the complexity of the PODE/CRN computation framework can be related
to that of conventional computation on a Turing machine, apart from the
characterization of the polynomial time complexity class \cite{BGP16icalp}.

\section{Conclusion}
\label{sec:conclusion}

In this paper, we have presented a CRN converter that allows us to read an analog input
such as the concentration of a chemical species by displaying its dyadic
representation as
a series of binary spikes on the concentration of an output species. We also conduct an
in-depth analysis of the precision that is theoretically possible to reach with
such a device according to the tuning of the three main types of rates that
appear in the reaction: the clock $\kc$, the approximate-majority $\km$ and the basic
operations $\ke$. In particular, we show that for fixed $\ke$ and
$\km$ adjusting the clock leads to a reading time that increases quadratically with the
desired precision while keeping all the concentration within constant bounds.

\vspace{1em}

We thus have contributed to both theoretical questions exposed in the
introduction.

Concerning the computation with PODE, we have shown that it is indeed possible
to construct a PODE that takes in input a bounded concentration and a desired
precision and returns the dyadic encoding of a number approaching the input with
the desired precision. Provided that we provide guaranty on the behavior of the
clock and the halter module on a formal level, potentially by reasoning on a
more complex CRN, and get to wrap everything together,
this would validate the computational framework proposed in
\cite{FLBP17cmsb} on a more practical level.


Regarding the Hartmanis-Stearns conjecture, this clarifies the picture. If it is
possible to compute the value of an algebraic irrational number in real time, it
is actually more difficult to read the output value once the computation has
reached a sufficient precision. Indeed, the
reader module is either bounded by a constant and quadratic in the precision
needed or linear in time but bounded by a variable that scales like the
precision. The difficult question to determine how these behaviors may be
related to the number of transitions of a multi-tape Turing machine is still
pending.


\vspace{1em}

This work could be extended in several directions. The most interesting one
would be to replace the clock with a handshaking mechanism. The idea would be to
wait until sufficient precision is reached before launching the next step of the
computation. This would allow for a kind of dynamical adjustment of the time spent
on the different cycles, going fast when the input is close to $0$ or $1$ and
taking more time when the comparison is difficult.

It is also unclear if this CRN is optimal in terms of time and space
complexity for the desired behavior. As this kind of analog framework is
relatively new, very few works exist to determine such bounds and even, what
are the relevant quantities to look at. In \cite{BGP16icalp}, Pouly proposes the length of
the solution curve and shows that it allows us to characterize the P-time class. But
how it is possible to build other classes of complexity inside the analog computation
framework and how to relate them to the ones of the logical complexity is still
a field largely unexplored.

Finally, we also start working at the other end of the chain: looking at how we can build
a concentration satisfying a desired precision from a dyadic encoding as input. As we try
to ensure some form of robustness with respect to the timing and intensity of
the spikes, this work is surprisingly more difficult than the one discussed
here. But it is in a good way and may be the subject of a short communication in
future work. 

\bibliographystyle{plain}
\bibliography{contraintes.bib}

\newpage
\appendix

\section{Appendix}

\subsection{Precision computation}
\label{app:precision}

Most differential equations involved in the dyadic converter harbor
converging exponential as solutions, that is, functions of the form:
$$f(t) = f_\infty + \left( f_0 - f_\infty \right) e^{-\ke t},$$
where $f_0$ is the value of the concentration at $t=0$, $f_\infty$ is the
desired steady state, and $\ke$ is the rate constant of the reaction. For model
(\ref{eq:full_crn}), this covers all the reactions for which the rate is set to
``rate''.

Thus, if we suppose that the reactions are active for a time larger than $\tau$,
the function is stopped at a value closer to $f_\infty$ than: 
$f_\infty + (f_0 - f_\infty) e^{-\ke \tau}$,
introducing the adimensional quantity $\epsilon = e^{-\ke \tau}$ we can bound
the final value $f_\tau$ by (in the case where $f_0 < f_\infty$):
\begin{equation}
    f_\tau \in [ f_0 + (f_0 - f_\infty) \epsilon, f_\infty ]
\end{equation}

Knowing this allows us to compute the bounds on the value of the different
species during the time evolution of one cycle of our converter module as
presented in table~\ref{tab:bounds}; and at the end,
the concentration $I_f$ of the Input species at the end of one clock cycle as a
function of its concentration $I_0$ at the start of the cycle,
specifically:
\begin{equation}
    \begin{cases}
        I_f \in [2I_0 - 3\epsilon, 2I_0] &\text{if } I_0 < \frac{1}{2}, \\
        I_f \in [2I_0 - 1 - 2\epsilon, 2I_0 - 1 + \epsilon] &\text{if } I_0 >
        \frac{1}{2}.
    \end{cases}
\end{equation}

\begin{table}
    \centering
    \begin{tabular}{l|ccccc}
           &   Input &  Tempo & Copy    &   Half    & Output \\
    Step 0 & $I_0$     & $0$    & $0$     & $0$       & $0$ \\
    Step 1 & $I_0$     & $[I_0(1-\epsilon), I_0]$ & - & $[\frac{1}{2}(1-\epsilon),
    \frac{1}{2}]$    & - \\
    Step 4 & $[I_0+I_0(1-\epsilon)^2, 2 I_0]$ &  $[0, I_0 (1 - \epsilon)]$ & 
    $[I_0(1-\epsilon)^2, I_0]$ & - & - \\
    Step 5 (O) & - & - & $[0, \frac{\epsilon}{\tau \ke}]$ & $\simeq I_0+\frac{1}{2}$ & - \\
    Step 5 (I) & - & - & $\simeq I_0+\frac{1}{2}$ & $[0, \frac{\epsilon}{\tau \ke}]$ & - \\
Step 6 (O) & - & - & - & - & $[0, \epsilon]$ \\
Step 6 (I) & - & - & - & - & $[1-\epsilon, 1]$ \\
Step 7 & - & $[0, \epsilon]$ & $[0, \epsilon]$ & $[0, \epsilon]$ & - \\
Step 8 (O) & $[I_0 + I_0(1-\epsilon)^2-\epsilon, 2I_0]$ & - & - & - & $[0, \epsilon]$ \\
Step 8 (I) & $[I_0 + I_0(1-\epsilon)^2-1, 2I_0-1+\epsilon]$ & - & - & - & $[0, \epsilon]$
    \end{tabular}
    \caption{Bounds on the concentration of the main species of the dyadic converter
    after the end of the different steps. When needed, we point to the cases
    where the output species have spiked (I) or not (O). Minus signs indicate
that the value does not have changed.} 
    \label{tab:bounds}
\end{table}

Considering the worst case with an error of order $3 \epsilon$ this means that at each step, the
value of the Input accumulated an error like: $e_{n+1} = 2 e_n +
3 \epsilon$, so that after $n$ steps, the error scales like:
$$e_n = 3 \left( 2^n - 1 \right) \epsilon,$$
thus if we still want good precision at step $p$, this means that we have to impose:
\begin{equation}
    \begin{aligned}
        3 \cdot 2^p \epsilon &\ll 1, \\
        e^{-\ke \tau} &\ll 2^{-p}, \\
        \ke \tau &\gg p.
    \end{aligned}
\end{equation}

Now, reaching step $p$ asks for a time $T(p) = 8 \tau p$ as the clock cycle
comports eight steps. As we have $\tau \propto p$ by our previous computation, we
can conclude that reaching a given precision asks for a time that scales quadratically:
\begin{equation}
    T(p) \propto \frac{p^2}{\ke}.
\end{equation}

\subsection{Convergence of approximate majority}
\label{app:AM}

Let us remind the ODEs associated to the model (\ref{CRN:AM}):
\begin{equation}
    \begin{aligned}
    \dot{A} &= -k_1 A H + k_2 U A \\
    \dot{H} &= -k_1 A H + k_2 U H \\
    \dot{U} &= 2 k_1 A H - k_2 U (A+H)
    \end{aligned}
\end{equation}

The obvious conserved quantity ($A+H+U = C_t$) allows us to introduce the
reduced variables: $a = \frac{A}{C_t}$, $u = \frac{U}{C_t}$ and $(1-a-u)$ as the
equivalent for $H$. We obtain the equations\footnote{To be exact, we should also
replace $k_1$ and $k_2$ by their rescaled version $\tilde{k} = k C_t$, we omit
this detail for the sake of readability.}:
\begin{equation}
    \begin{aligned}
        \dot{a} &= - k_1 (a (1-a-u)) + k_2 u a \quad &a(t=0) = \alpha\\
    \dot{u} &= 2 k_1 (a (1-a-u)) - k_2 u (1-u) \quad &u(t=0) = 0
    \end{aligned}
\end{equation}

Once again a change of variable is necessary to simplify these equations, we use
the variable $\tilde{a} = a + \frac{u}{2}$, giving:
\begin{equation}
    \begin{aligned}
    \dot{ \ta} &= k_2 u (\ta - \frac{1}{2}) \\
    \dot{u} &= 2 k_1 (\ta - \frac{u}{2}) (1-\ta-\frac{u}{2}) - k_2 u (1-u)
    \end{aligned}
\end{equation}

Supposing that $k_1 \gg k_2$ allows us to approximate $u$ as a fast variable with
the equilibrium given by:
\begin{equation}
    u = \left( \sqrt{4 \ta (1-\ta) + 1} - 1\right),
\end{equation}
that we can plug in the equation of $\ta$. As this equation is separable, we
obtain the relation:

\begin{equation}
    \begin{split}
        k_2 t + c_1 = -\log(\sqrt{s} - 1) - log(\sqrt{s} - 2 \ta - 1) + (1 +
        \sqrt{2}) \log(\sqrt{s} - 2 \sqrt{2} \ta + 2 \ta - 1) \\ - (\sqrt{2} - 1)
        \log(\sqrt{s} + 2 (1 + \sqrt{s}) \ta - 1) = f_i(\ta)
    \end{split}
\end{equation}
where for readability, we denote: $\sqrt{4 \ta (1-\ta) +1} = \sqrt{s}$ and
$c_1$ is an integration constant that should be computed
from the initial condition. The function $f_i$ is symetrical under the relation
$\ta \rightarrow 1-\ta$ and describes the evolution of the system over time.
That is, if we want a final value $r$ while starting at $\ta_0$, we have to wait
for a time: $t = \frac{f_i(r) - f_i(\ta_0)}{k_2}$. In our case, we have a fixed
time $\tau$ a precision requirement $r$ and want to determine how close we can
be to one half to still reach it. That is, we want to know $\delta$ such that:
$$k_2 \tau = f_i(r) - f_i(\frac{1}{2}-\delta).$$

An expansion around both $r \ll 1$ and $\delta \ll 1$ gives us:
\begin{equation}\begin{aligned}
    k_2 \tau &= -\log 4 r^2 - (1+\sqrt{2}) \log \delta \\
    \log \delta &= -\frac{\log 4 r^2}{1+\sqrt{2}} - \frac{k_2 \tau}{1+\sqrt{2}}
\end{aligned}\end{equation}

From this, we can estimate the scaling of the critical threshold, below which the
approximate majority is not able to discern the majority species as a function of
$k_2$ and $\tau$:
\begin{equation}
    \delta^\star \propto \exp{-\frac{k_2 \tau}{1+\sqrt{2}}}.
\end{equation}

\vspace{1em}

All this derivation has been done under the assumption that $k_1 \gg k_2$ which
is the most natural one. To be complete, suppose that $k_1 \ll k_2$. We can
easily prove that, once again, $u$ equilibrates quickly near $0$ giving us: $u = \frac{2
k_1}{k_2} \ta (1-\ta)$ and inserting it in the equation over $\ta$ we have:
\begin{equation}
    \dot{ \ta} = 2 k_1 \ta (1-\ta) (\ta - \frac{1}{2}).
\end{equation}

In this simpler case, we have the analytical solution:
\begin{equation}
    \ta(t) = \frac{4 e^{c_0} + e^{k_1 t} \pm \sqrt{4 e^{c_0+k_1 t}+e^{2k_1 t}}}
                  {2(4e^{c_0}+e^{k_1 t})},
\end{equation}
where $c_0$ is the integration constant that can be derived from the initial
condition. The two important elements to remark are that the $\pm$
sign depends on whether $\alpha$ is initially higher or lower than one half and that
the system follows the same kind of sigmoidal pattern as previously, starting by a slow phase
to escape the intermediate region followed by an exponential convergence toward
one of the two attractors that are $0$ and $1$.

From this, we can derive the same kind of relation as previously:
$$\delta^\star \propto e^{-\frac{k_1 \tau}{2}}.$$

\subsection{Integrating over a cosine}
\label{app:cosine}

We approximate the sine wave of our clock by square signal to ease the analysis.
Actually, we can pretty easily compute the equivalent of our $\epsilon =
\exp{- \tau \ke}$ quantity for a sinusoidal wave of frequency $\omega$.

Let us use the activator species $T$ to activate two reactions, thus creating a
simple steady state for the molecule $R$ (for reporter):
$T \xleftrightarrow{\ke} T+R$
and use a sinusoidal profile:
\begin{equation}
    T(t) = \begin{cases}
        \sin{t \omega \pi} &\text{if } t \in [ 0, \omega^{-1}] \\
        0 &\text{otherwise}
    \end{cases}
\end{equation}

The associated ODE is: $\dot{A} = \ke T (1-A)$ that is separable, and thus we
easily obtain:
\begin{equation}
    A_\infty = 1 - e^{- \frac{2 \ke}{\omega \pi}} = 1 - \epsilon.
\end{equation}

Hence, in our case where $\omega = \pi^{-1}$, we simply have $\epsilon = e^{-2
\ke}$ where $\ke$ is the rate constant of the main reaction of the current step.
Or said otherwise, the sine wave is equivalent to a square signal of duration
$\tau = 2$.

\subsection{Linker CRN}
\label{app:linker}

The following CRN presents a prototype of linker CRN that allows to use our
reader CRN on the output of a computing CRN as presented by Fages et al.  Here,
$y$ is the species mimicking the precision of the main CRN, $a$ is an activator
that spikes when the desired precision $p$ is reached.  Finally,
$s_\text{compute}$ and $s_\text{read}$ represent the two species of the switch
that turns the whole CRN from computing to reading mode by catalyzing every
reactions in their respective parts.

\begin{small}
\begin{equation}
\begin{aligned}
    y+s_\text{compute} &\xrightarrow{1} s_\text{compute}, &\quad
    a &\leftrightarrow \emptyset, \\
    y+a+p &\xrightarrow{2 k_f} y+2 \cdot a+p, &\quad
    2 \cdot y+a &\xrightarrow{k_f} 2 \cdot y, \\
    2 \cdot p+a &\xrightarrow{k_f} 2 \cdot p, \\
    s_\text{compute} + a &\xrightarrow{1} s_\text{read} + a, &\quad
    s_\text{compute} + s_\text{read} &\xrightarrow{1} 2 \cdot s_\text{read}.
\end{aligned}
\end{equation}
\end{small}
With at $t=0$, $s_\text{compute}=1$ and $a=1$.

\end{document}